\documentclass[aps,prb,amsfonts,amssymb,twocolumn,amsmath,preprintnumbers,nofootinbib,floatfix,showpacs]{revtex4}
\usepackage{amsmath,bm,amsfonts,amssymb}
\usepackage[dvips]{graphics}
\usepackage{graphicx}
\usepackage{bm}

\begin{document}

\title{Nonequilibrium Transport in Superconductor/Ferromagnet/Superconductor Diffusive
Junctions: Interplay between Proximity Effect and Ferromagnetism.}
\author{I. V. Bobkova}
\affiliation{Institute of Solid State Physics, Chernogolovka,
Moscow reg., 142432 Russia}
\author{A. M. Bobkov}
\affiliation{Institute of Solid State Physics, Chernogolovka,
Moscow reg., 142432 Russia}

\date{\today}

\begin{abstract}
The theory of the I-V characteristics in diffusive
superconductor/weak ferromagnet/superconductor (SFS) junction is
developed. We show that the exchange field $h$ of the ferromagnet
manifests itself as an additional conductance peak at $eV \sim \Delta+h$ in
the phase-coherent regime, when the Thouless energy is of the
order of superconducting order parameter. The excess current
exhibits non-monotonous dependence on the exchange field and
non-trivial temperature behavior, which is strongly influenced by
the temperature dependence of the exchange field.
\end{abstract}

\maketitle

The recent progress in the experimental techniques has made
possible the fabrication of mesoscopic structures on the nanometer
scale. Hybrid structures containing superconducting and
ferromagnetic elements offer an opportunity to generate and
control coherent spin transport with otherwise conventional
electronics. The equilibrium transport and proximity effect in
such structures have been theoretically and experimentally
investigated recently in details as for the case of weak
ferromagnetic alloys so as for half-metals like $CrO_2$
\cite{Eschrig03,Keizer06}.In particular, Josephson current  in SFS
junctions and $T_c$ of SF bilayers and multilayers have been
investigated in details for the case of weak ferromagnetic alloys
(see Ref.\onlinecite{buzdin} and references therein). Equilibrium
density of states was also studied \cite{kadigrob,Fazio99}. On the
other hand, to the best of our knowledge, the nonequilibrium
transport in SFS mesoscopic junctions has not been studied yet
neither from theoretical nor from experimental point of view
except for theoretical investigations of magnetic quantum point
contacts\cite{Cuevas01,Fogelstrom02,Bobkova06,Sauls06}. We address
our theoretical paper to the part of this problem and study the
phase-coherent transport in diffusive voltage-biased SFS plane
junctions. We consider interlayers made of a weak ferromagnetic
material, in which the value of the exchange field $h$ (measured
in the energy units) is of order of superconducting order
parameter $\Delta$. In ferromagnetic alloys like $CuNi$, which
have been intensively used by now for experimental investigation
of equilibrium properties of SFS heterostructures, the exchange
field is several times larger than $\Delta$. On the other hand, it
can be concluded from our analysis of the problem that for
studying the I-V characteristics of SFS junctions, made on the
basis of a weak ferromagnetic alloy, the most interesting case is
$h \lesssim \Delta$. As far as we know, the work on the creation
of appropriate alloys is in progress now, so we believe that this
limit can be experimentally realized in the nearest future.

The I-V characteristics of the superconductor/normal
metal/superconductor (SNS) voltage-biased diffusive junction are
studied theoretically in details mainly in two limits. The first
one is the limit of short junction $d \ll \xi$, where $d$ is the
length of the interlayer, $\xi=\sqrt{\hbar D/\Delta}$ is the
superconducting coherence length and $D$ is the diffusion
constant. In this regime the subharmonic gap structure (SGS) in
the differential conductance $dI/dV$ consists of a set of
pronounced maxima at $eV_n=2\Delta/n$
\cite{Bardas97,Kupriyanov03}. The excess current, which takes
place at high biases $eV \gg \Delta$, behaves in dependence on
temperature as $\Delta(T)$ does and can be as positive so as
negative depending on the transparency of SN
interfaces\cite{Volkov93,Bardas97}. The second one is the
incoherent limit $d \gg \xi$. The proximity effect is negligible
in this regime and the transport is determined by the kinetic
equation for the distribution
function\cite{Klapwijk82,Octavio8388,Bezuglyi00}. The SGS of the
conductance again shows sharp features at $eV_n=2\Delta/n$. At the
same time in the intermediate regime $\xi \sim d$ the proximity
effect, which manifests itself as a minigap $\Delta_g$ in the
equilibrium density of states in the normal region, takes place.
It has been shown very recently \cite{Cuevas05} that for SNS
voltage-biased diffusive junction with highly-transparent NS
interfaces in this regime, when the interplay between proximity
effect and MARs takes place, the well-known subgap conductance
structure $eV_n=2\Delta/n$ modifies and exhibits an additional
maximum at roughly $eV \sim \Delta + \Delta_g$. These predictions
are in good agreement with the existing
experiments\cite{Kutchinsky97,Hoss00}. In the present work we
numerically calculated the I-V characteristics taking into account
the proximity effect and non-equilibrium distribution function in
the ferromagnetic interlayer. We show that in the phase-coherent
regime, when the proximity effect in the ferromagnetic region
takes place, the exchange field of a weak ferromagnet $h<\Delta$
explicitly manifests itself in the SGS of diffusive voltage-biased
SFS junction leading to the additional maximum at $eV \sim (\Delta
+ h)$, which is the most pronounced in the case of low-transparent
SF boundaries. For the case $h>\Delta$ the differential
conductance exhibits no pronounced characteristic features. The dc
current at high voltage biases is also investigated. It is deficit
($I_{def} = I-V/R<0$) for the case of highly-resistive interfaces
we consider and has a non-monotonous behavior as a function of
exchange field, reaching the maximum value at roughly $h \sim
\Delta$. This non-monotonous behavior results in the fact, that
the temperature dependence of the deficit current exhibits the
characteristic features, which are determined by the temperature
dependence of the exchange field and can be used for its
experimental mapping.

Further the model under consideration and the method we use are
briefly described. We study an SFS junction, where F is a
diffusive weak ferromagnet of length $d$ coupled to two identical
superconducting (S) reservoirs. The superconductors are supposed
to be diffusive and have $s$-wave pairing. We assume the SF
interfaces to be not fully transparent and suppose that the
resistance of the SF boundary $R_g$ dominates the resistance of
the ferromagnetic interlayer $R_F$. We use the quasiclassical
theory of superconductivity for diffusive systems in terms of
time-dependent Usadel equations \cite{usadel}. The fundamental
quantity for diffusive transport is the momentum average of the
quasiclassical Green's function $\check g(x,\varepsilon, t) =
\langle \check g(\bm p_f, x,\varepsilon, t) \rangle_{\bm p_f}$. It
is a $8\times8$ matrix form in the product space of Keldysh,
particle-hole and spin variables. Here $x$ - is the coordinate
measured along the normal to the junction, $t$ stands for a time
variable and $\varepsilon$ is the excitation energy.

The electric current should be calculated via Keldysh part of the
quasiclassical Green's function. For the plane diffusive junction
the corresponding expression reads as follows
\begin{widetext}
\begin{equation}
\frac{j^{el}}{e} = -\frac{d}{8 \pi^2 e^2 R_F} \int
\limits_{-\infty}^{+\infty} d \varepsilon {\rm Tr}_4
\left[\frac{1}{2}(\hat \tau_0 + \hat \tau_3) \hat \sigma_{0}
\left(\check g(x, \varepsilon, t)\otimes \frac{\partial \check
g(x, \varepsilon, t)}{\partial x}\right)^K \right] \label{current}
\enspace ,
\end{equation}
\end{widetext}
where $e$ is the electron charge and $\hbar = 1$ throughout the
paper. $\left(\check g(x, \varepsilon, t)\otimes (\partial \check
g(x, \varepsilon, t))/(\partial x)\right)^K$ is a $4\times4$
Keldysh part of the corresponding combination of full Green's
functions. The product $\otimes$ of two functions of energy and
time is defined by the noncommutative convolution $A \otimes B =
e^{i(\partial_\varepsilon^A\partial_t^B-\partial_t^A\partial_\varepsilon^B)}A(\varepsilon,t)B(\varepsilon,t)$.
$\hat \tau_i$ and $\hat \sigma_i$ are Pauli matrices in
particle-hole and spin spaces respectively.

The quasiclassical Green's function $\check g(x,\varepsilon, t)$
satisfies the non-stationary Usadel equation. In order to solve
the Usadel equation it is convenient to express quasiclassical
Green's function $\check g$ in terms of Riccati coherence
functions $\hat \gamma^{R,A}$ and $\hat {\tilde \gamma}^{R,A}$ and
distribution functions $\hat x^K$ and $\hat {\tilde x}^K$. All
these functions are $2 \times 2$ matrices in spin space and depend
on $(x, \varepsilon, t)$. The corresponding expression for $\check
g$ can be found in Ref.\onlinecite{Eschrig00} and therefore is not
explicitly written here. Riccati coherence and distribution
functions obey Riccati-type transport equations\cite{Eschrig04},
where the appropriate self energy takes the form $\check \Sigma(x,
\varepsilon, t) = h \hat \sigma_3 + (1/2\pi \tau_s) \hat \sigma_3
\check g \hat \sigma_3$. Here $\tau_s^{-1}$ is an inverse magnetic
scattering time. As the F layer is supposed to be an alloy, a role
of magnetic scattering may be quite
important\cite{Sellier03,Ryazanov04}. We assume a presence of the
relatively strong uniaxial magnetic anisotropy which prevents
mixing of spin-up and spin-down Green's functions, so the magnetic
scattering term is a diagonal matrix in spin
space\cite{Ryazanov05}.

The Riccati-type transport equations should be solved together
with the boundary conditions at SF interfaces. As it was mentioned
above we consider the case when the dimensionless conductance of
the boundary $G \equiv R_F/R_g \lesssim 1$, so the interface
transparency $T \sim G(l/d) \ll 1$ ($l$ is the mean free path).
Due to the smallness of the interface transparency $T$ we can use
Kupriyanov-Lukichev boundary conditions at SF
boundaries\cite{Kupriyanov}. In terms of Riccati coherence and
distribution functions they take the form
\begin{widetext}
\begin{equation}
2 i \pi d \partial_x \hat \gamma^R_{l,r} = \pm G \left[ \hat
f^R_{S;l,r} + \hat \gamma^R_{l,r} \otimes \hat {\tilde
g}^R_{S;l,r} - (\hat g^R_{S;l,r} + \hat \gamma^R_{l,r} \otimes
\hat {\tilde f}^R_{S;l,r}) \otimes \hat \gamma^R_{l,r} \right]
\enspace , \label{boundary_cond_gamma}
\end{equation}
\begin{equation}
2 i \pi d \partial_x \hat x^K_{l,r} = \pm G \left[ (1/2) \left(
\hat g^K_{S;l,r} + \hat \gamma^R_{l,r} \otimes \hat {\tilde
g}^K_{S;l,r} \otimes \hat {\tilde \gamma}^A_{l,r} \right) - \hat
\gamma^R_{l,r} \otimes \hat {\tilde f}^K_{S;l,r} - \left( \hat
g^R_{S;l,r} + \hat \gamma^R_{l,r} \otimes \hat {\tilde
f}^R_{S;l,r} \right) \otimes \hat x^K_{l,r} - h.c. \right]
\enspace . \label{boundary_cond_x}
\end{equation}
\end{widetext}
Here Riccati coherence and distribution functions denoted by the
lower case symbols $l,r$ are taken at the left and right ends of
the ferromagnet. The quantities denoted by the lower case symbols
($S;l,r$) are corresponding Green's functions at the
superconducting side of the left and right SF interfaces. We
assume the parameter $(R_F/R_g)(\sigma_F/\sigma_s)$, where
$\sigma_F$ and $\sigma_s$ stand for conductivities of
ferromagnetic and superconducting materials respectively, to be
also small, what allows us to neglect the suppression of the
superconducting order parameter in the S leads near the interface
and take the Green's functions at the superconducting side of the
boundaries to be equal to their bulk values with the appropriate
shift of the quasiparticle energy due to the applied voltage
$V=V_r-V_l$.

Let us now turn to the analysis of the I-V characteristics.

\begin{figure}[!tbh]
   \centerline{\includegraphics[clip=true,width=3in]{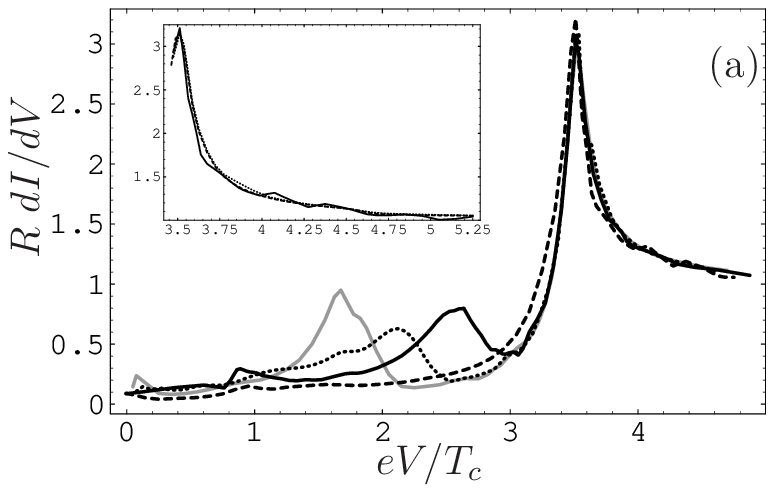}}
   \centerline{\includegraphics[clip=true,width=3in]{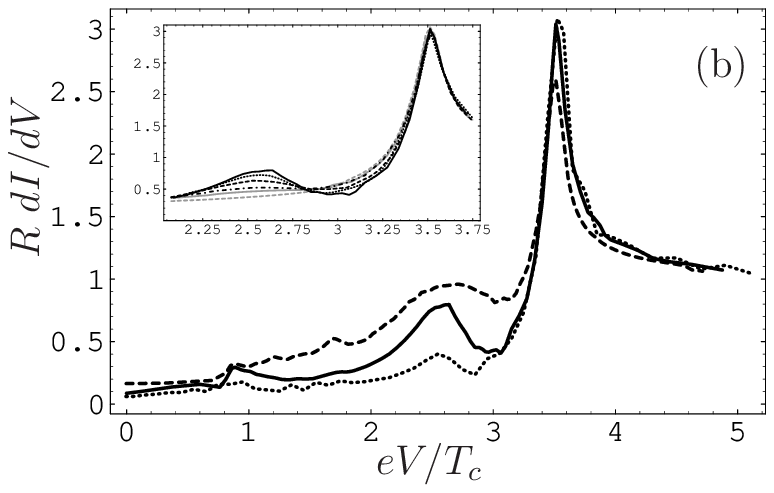}}
   \caption{The zero temperature differential conductance $dI/dV$,
normalized to its value for the normal state of the leads, as a
function of $eV/T_c$, taken for: (a) different exchange fields:
$h=0$(gray solid); $h=0.45T_c$(dotted); $h=0.9T_c$(black solid)
and $h=2.6T_c$(dashed); The insert to (a) shows the $R dI/dV$ as a
function of $eV/T_c$ for $h>\Delta$: $h=2.6T_c$(solid);
$h=3T_c$(dashed) and $h=3.5T_c$(dotted) are practically
undistinguishable; (b) different values of the Thouless energy:
$\varepsilon_{Th}=5T_c$(dashed); $\varepsilon_{Th}=2T_c$(black
solid) and $\varepsilon_{Th}=0.5T_c$(dotted). $T_c$ is the
superconducting critical temperature. $G=0.1$ and $1/\tau_s=0$.
The inset to (b) demonstrates the influence of the magnetic
scattering on the conductance: $1/\tau_s=0$ (solid);
$1/\tau_s=0.05 T_c$(dotted); $1/\tau_s=0.15 T_c$(dashed);
$1/\tau_s=0.3T_c$ (dashed-dotted); $1/\tau_s=0.45T_c$ (gray
solid); $1/\tau_s=0.9T_c$ (gray dashed). $h=0.9 T_c$.}\label{IV}
\end{figure}

Figure \ref{IV}(a) shows the dependence of the zero temperature
differential conductance $dI/dV$ on voltage for different values
of the exchange field in the ferromagnet. The following main
features can be observed:

(i) there is a series of peaks at $eV_n=2\Delta/n$. In fact, only
the peaks corresponding to $n=1,2$ are seen in the figure. The
other harmonics are smeared out due to the small interface
transparency and large enough intrinsic broadening of the
quasiparticle energy (modelling the inelastic scattering rate)
$\gamma=0.03 T_c$, which we take in the numerical calculations.
This value of intrinsic broadening is large, but consistent with
an estimate of inelastic scattering rate due to electron-phonon
processes $\gamma \sim \varepsilon^3/\omega_D^2$ at high enough
energies $\varepsilon \sim \Delta$ \cite{gantmakher}.

(ii) for $h \lesssim \Delta$ the large additional maximum appears
at roughly $eV=\Delta+h$, while the curves corresponding to
$h>\Delta$ do not exhibit any additional pronounced characteristic
features, in particular at $eV=\Delta+h$ (See Insert to
Fig.\ref{IV}(a)). The exchange fields of order of several $\Delta$
only result in the suppression of the peak at $eV=\Delta$. In
fact, the larger the exchange field the less different from each
other the appropriate conductance curves. Of course, this
statement is related to the region of weak exchange fields we
study.

Analyzing the dependence of the I-V characteristics on the
Thouless energy $\varepsilon_{Th}=D/d^2$, which is represented in
Figure \ref{IV}(b), we can conclude, that the most pronounced
maximum at $eV \sim \Delta+h$ exists at the region
$\varepsilon_{Th} \sim \Delta$. If the Thouless energy increases
the peak washes out and finally vanishes. At the limit of very
short junction $\varepsilon_{Th} \gg \Delta$ the only structure
$eV=2\Delta/n$ survives. On the other hand, with decreasing the
Thouless energy the amplitude of maximum at $eV \sim \Delta+h$
reduces to the zero value at the limit of very small Thouless
energies and again the only structure at $eV=2\Delta/n$ can be
observed. With the increasing of transparency of the interfaces
the maximum smears out and is masked by the growing peak at
$eV=\Delta$.

The inset to Figure \ref{IV}(b) demonstrates the influence of the
magnetic scattering on the maxima discussed above. It can be seen
that the magnetic scattering does not influence the peak at
$eV=2\Delta$, while the maximum at $eV \sim \Delta + h$ is
broadened, but not dramatically for small compared with $T_c$
values of magnetic scattering rate. Of course, large enough values
of $1/\tau_s \sim h,T_c$ destroy the peak and we found the
characteristic value of $1/\tau_s$, under which the peak
completely washes out, to be approximately equal to $0.45T_c$. The
magnetic scattering is typically arises due to the magnetic
inhomogeneity, related above all to $Ni$-rich clusters, and can be
of order of the average exchange field in the ferromagnetic alloys
like $Cu_{1-x}Ni_x$ (see \onlinecite{Ryazanov05} and references
therein). However, one can believe that the magnetic inhomogeneity
could be weaker in the more diluted alloys with very low exchange
fields of order of $\Delta$ resulting in small values of magnetic
scattering parameter $1/\tau_s$, even compared with $T_c$.

The appearance of this additional maximum in the intermediate
regime $\varepsilon_{Th} \sim \Delta$ is caused by the proximity
effect in the ferromagnetic region and can be qualitatively
understood as follows. It is well known that for a diffusive SNS
junction the zero-bias density of states (DOS) has a minigap in
the normal region. For low-transparency junctions it is $\Delta_g
\propto G\varepsilon_{Th}$. If the normal metal is replaced by a
ferromagnet, the exchange field $h$ shifts the densities of states
for the two spin subbands in the opposite directions, therefore
the minigap in the spectrum closes at $h \sim \Delta_g$
\cite{Fazio99}, but the sharp onsets in the DOS for the both
subbands survive until they merge the edge of the continuous
spectrum. So for small enough values of $\Delta_g < h$ and $eV
\sim \Delta + h \pm \Delta_g$ an electron belonging to the one
spin subband travels through the interlayer from one sharp onset
of the DOS at $-\Delta$ to another one at $h \pm \Delta_g$, while
an electron from the other spin subband can move from $-h \pm
\Delta_g$ to the edge of continuous spectrum at $+\Delta$.
According to this qualitative consideration the maximum should be
split and the splitting is roughly proportional to the value of
the equilibrium minigap $\Delta_g \sim G \varepsilon_{Th}$. We do
not observe such splitting at our curves. In addition, due to the
multiple Andreev reflection processes there could arise the
subsequent harmonics at $eV \sim (\Delta+h\pm\Delta_g)/n$. Of
course, the proposed qualitative explanation is very crude because
it deals with the equilibrium minigap instead of real
position-dependent pseudogap at finite bias and does not take into
account the behavior of the distribution function in the
interlayer, which is highly nonequilibrium and obviously
influences the I-V characteristics.

\begin{figure}[!tbh]
   \centerline{\includegraphics[clip=true,width=3in]{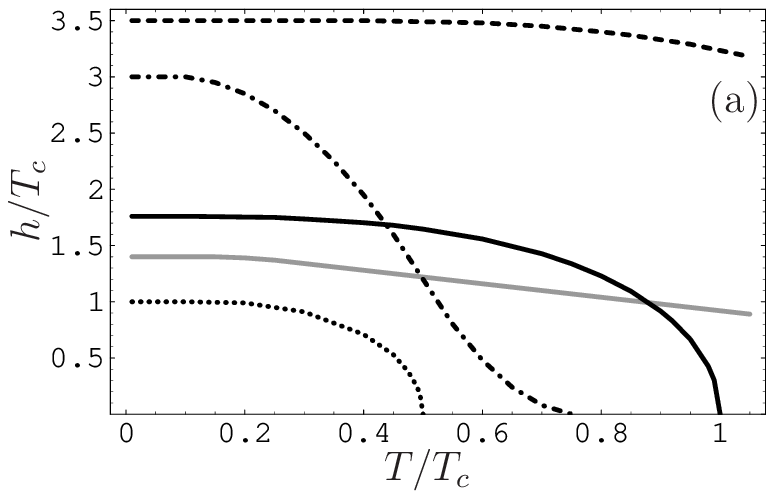}}
             \centerline{\includegraphics[clip=true,width=3in]{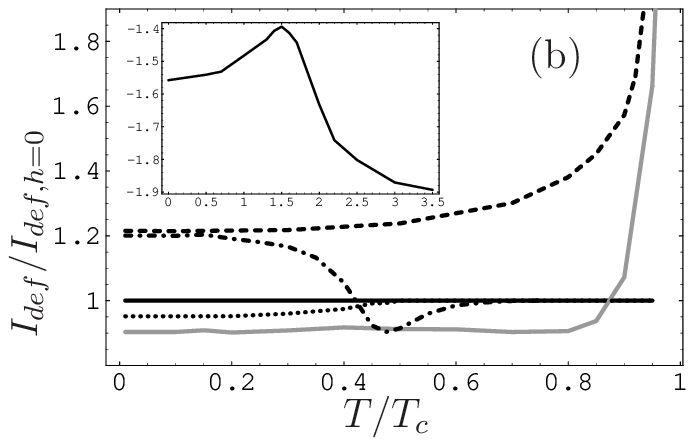}}
       \caption{(a) Possible dependencies of the weak exchange field on temperature.
  The black solid line represents the BCS-type behavior of $\Delta(T)/T_c$. (b) The appropriate temperature
  dependencies of the deficit current, normalized to its value at
  $h=0$. The inset shows the deficit current $eIR/T_c$ as a function of the
  exchange field at $T=0$, $G=0.1$, $\varepsilon_{Th} = 2 T_c$. \label{exc_cur}}
\end{figure}

Let us now discuss the behavior of the high voltage $eV \gg
\Delta$ deficit current $I_{def} = I-V/R<0$, where $R=R_F + 2 R_g$
is the full resistance of the junction. The inset in Figure
\ref{exc_cur} represents the typical dependence of the deficit
current on the exchange field. The deficit current is a
non-monotonous function of the exchange field and reaches a
maximum at $h \sim \Delta$. The non-monotonous behavior of the
deficit current leads to very peculiar dependencies of this
current on temperature. The corresponding curves have the
characteristic features determined by the temperature dependence
of the exchange field. Figure \ref{exc_cur}(a) shows several
examples of possible behavior of exchange field on temperature in
comparison with the temperature dependence of $\Delta(T)$. Figure
\ref{exc_cur}(b) represents the corresponding dependencies of the
deficit current on temperature. They are normalized to the value
of the deficit current at zero exchange field. It is seen that (i)
if $h(T)<\Delta(T)$ in the whole temperature region $0<T<T_c$, the
curve $\tilde I_{def}(T) = I_{def}(T)/I_{def,h=0}(T)$ resembles
the appropriate dependence $h(T)$ up to the Curie temperature of
the ferromagnet $T_{Cu}$ and then follows the curve for $h=0$
(dotted curve), (ii) in the opposite case of $h(T)>\Delta(T)$ for
$0<T<T_c$ $\tilde I_{def}(T)$ monotonously increases up to $T=T_c$
(dashed curve), (iii) if $h(0)>\Delta(0)$ and $T_{Cu}<T_c$, then
$\tilde I_{def}(T)$ has a minimum approximately at the
intersection point of the curves $h(T)$ and $\Delta(T)$. For the
temperatures higher then $T_{Cu}$ it follows the curve for $h=0$
(dashed-dotted curve) and (iv) if $h(0)<\Delta(0)$ and
$T_{Cu}>T_c$, then $\tilde I_{def}(T)$ grows starting from the
values less then that one for $h=0$, intersects this line and
rises very sharply at $T \to T_c$ (gray solid curve).

In summary, we have developed the theory of the I-V
characteristics of diffusive SFS junctions with highly enough
resistive SF interfaces. It is found that weak exchange field of
the ferromagnet manifests itself only in the coherent regime
$\varepsilon_{Th} \sim \Delta$. For $h<\Delta$ it results in the
appearance of the additional well-pronounced maximum in the
differential conductance at $eV \sim (\Delta+h)$ if the magnetic
scattering rate in the ferromagnetic region is small compared to
$T_c$. We have also studied the deficit current at high voltages
and found that it behaves non-monotonously as a function of the
exchange field exhibiting a maximum at $h \sim \Delta$. This gives
rise to the characteristic features in the temperature behavior of
the deficit current, determined by the temperature dependence of
the exchange field. We believe, that all the discussed features
can be measured in the diffusive SFS junctions, realized on the
basis of weak ferromagnetic alloys.

We thank V.V. Ryazanov and A. Yu. Rusanov for discussions. The
support by RFBR Grants 05-02-17175 (I.V.B. and A.M.B.),
05-02-17731 (A.M.B.) and the programs of Physical Science Division
of RAS is acknowledged. I.V.B. was also supported by the Russian
Science Support Foundation.

\end{document}